\documentclass[sn-mathphys,Numbered]{sn-jnl}

\usepackage{graphicx}%
\usepackage{multirow}%
\usepackage{amsmath,amssymb,amsfonts}%
\usepackage{amsthm}%
\usepackage{mathrsfs}%
\usepackage[title]{appendix}%
\usepackage{xcolor}%
\usepackage{textcomp}%
\usepackage{manyfoot}%
\usepackage{booktabs}%
\usepackage{algorithm}%
\usepackage{algorithmicx}%
\usepackage{algpseudocode}%
\usepackage{listings}%

\begin{document}

\title[Schr{\"o}dinger cat state formation in small bosonic Josephson junctions at finite temperatures and dissipation]{Schr{\"o}dinger cat state formation in small bosonic Josephson junctions at finite temperatures and dissipation}

\author*[1]{\fnm{D.V.}\sur{Tsarev}}\email{dmitriy\_93@mail.ru}
\affil[1]{\orgdiv{Institute of Advanced Data Transfer}, \orgname{ITMO University}, \orgaddress{\street{Kronverksky Pr. 49, bldg. A,}, \city{St. Petersburg}, \postcode{197101}, \country{Russia}}}
\equalcont{These authors contributed equally to this work.}

\author[1]{\fnm{D.V.} \sur{Ansimov}}

\equalcont{These authors contributed equally to this work.}

\author[2]{\fnm{S.A.} \sur{Podoshvedov}}\email{dmitriy\_93@mail.ru}
\affil[2]{\orgdiv{Quantum Light Engineering Laboratory, Institute of Natural and Exact Sciences}, \orgname{South Ural State
University}, \orgaddress{\street{Lenin Av., 76}, \city{Chelyabinsk}, \postcode{454080}, \country{Russia}}}

\equalcont{These authors contributed equally to this work.}

\author[1,2]{\fnm{A.P.} \sur{Alodjants}}

\equalcont{These authors contributed equally to this work.}


\abstract
{
In this work, we consider the feasibility of Schr{\"o}dinger cat (SC) and $N00N$ states formation by a convenient bosonic Josephson junction (BJJ) system in two-mode approximation. Starting with purely quantum description of two-mode Bose-Einstein condensate we investigate the effective potential approach that provides an accurate analytical description for the system with a large number of particles. We show that in the the zero temperature limit SC states result from a quantum phase transition that occurs when the nonlinear strength becomes comparable with the Josephson coupling parameter. The Wigner function approach demonstrates the growth of the SC state halves separation and formation of $N00N$-like states (a Fock state superposition) with the particle number increase. We examine the possibility to attain the SC state at finite temperatures and a weak dissipation leading to appearing of some critical temperature; it defines the second-order phase transition from classical activation process to the SC state formation through the quantum tunneling phenomenon. Numerical estimations demonstrate that the critical temperature is sufficiently below the temperature of atomic condensation. The results obtained may be useful for experimental observation of SC states with small condensate Josephson junctions. 
}

%
%
%
%
%

\maketitle

\section{Introduction}

For more than 80 years, studies of the Schr{\"o}dinger cat (SC) states originated from the famous work by Erwin Schr{\"o}dinger~\cite{Sch} have been in the focus of scientists in quantum physics and beyond, see e.g.~\cite{Wheeler,Leggett,Dodonov,Gerry,Haroche}. The SC state represents a macroscopic superposition of two well-distinguished quantum states, which is at the heart of modern quantum theory. In quantum physics SC states are proposed to obtain based on optical~\cite{Ourjoumtsev,Sychev}, superconductor~\cite{Grimm}, semiconductor~\cite{Cosacchi}, and atomic~\cite{Johnson,Kai,Hacker} devices. Practically, SC states containing macro- or mesoscopically large numbers of particles may be crucial for fault-tolerant quantum computing~\cite{Eaton} and quantum metrology~\cite{Tatsuta}. Notably, so-called $N00N$ states represent an indispensable tool in some metrological applications at the Heisenberg level of sensitivity and may be considered as a limiting case of the SC state, cf.~\cite{Tsarev2018,Tsarev2019,Alodjants2022}. However, creation and manipulation by large amplitude SC states represent a non-trivial and challenging task for current quantum technologies both in theory and experiment, cf.~\cite{Mikheev,Takase,Tsarev2020}. In this regard, the effect of thermal fluctuations and losses is still an open and important problem in the formation of large amplitude SC states, see e.g.~\cite{Pawłowski}. 

In this work, we examine this problem for so-called bosonic Josephson junction (BJJ) systems, which represent two weakly linked oscillators with Kerr-like nonlinearity. Practically, such systems are obtained with tunnel-coupled waveguides~\cite{Jin}, two-mode Bose-Einstein condensates~\cite{Albiez, Vretenar}, and small superconducting Josephson junction devices~\cite{Devoret}. 

Currently, various BJJ systems represent a universal platform for the formation of SC state ~\cite{Cirac1998,Mazzarella2011,Piazza,Haigh2010} and maximally-path-entangled $N00N$ state~\cite{Bollinger1996,Dowling2000,Dowling2004,Dowling2008,Kulik}. Noteworthy, in the framework of BJJ devices it is possible to consider two possibilities for SC state formation. First, we can recognize SC states as a superposition of phase states, which occur in Kerr-like medium at some specific time domains, see~\cite{Piazza} and cf.~\cite{Yurke}. Another possibility examined here is based on some peculiarities of stationary superposition states, which appear in BJJ systems,~\cite{Tsarev2020,Cirac1998,Mazzarella2011,Haigh2010}. Obviously, in real-life experiments the finite temperatures as well as dissipative effects play an important role is such systems; highly nonclassical states discussed in this work are extremely sensitive to dissipation and decoherence depending on thermal fluctuations. 

How low should the temperature in a real-world experiment be for SC states to be obtained? It is evident that the answer to this question strictly depends on peculiarities of a given BJJ system. However, in some cases a satisfactory answer can be given by using simple models and applying some specific approaches. 
 
In this work, we propose an analytical model of SC and $N00N$ states formation, which enables to find critical parameters of the BJJ system coupled with the environment and relevant to the second-order phase transition from the quantum to classical behaviour, cf.~\cite{Chudnovsky1997}. To be more specific, we examine small atomic condensates with negative scattering length in two-mode approximation~\cite{Anglin}. For our results, we also assume that the BJJ is capable of weak dissipation. Thus, the purpose of this work is to specify physical conditions, under which SC and $N00N$ states can be obtained in atomic BJJ systems at finite temperatures in the presence of dissipation. 

In Sec.~2 we describe our BJJ model and identify conditions for SC state formation in the zero temperature limit. In Sec.~3 we develop an effective-potential method that helps derive an effective Schr{\"o}dinger equation for the condensate wave function given in the particle-number-state representation, cf. e.g.~\cite{Zaslavskii}. Then, to study thermal properties of the BJJ, we establish an approach based on the inverted potential analysis. This approach is commonly used to study non-zero-temperature transitions in various bistable spin systems, such as Lipkin model~\cite{Heiss1988,Glick1965}, magnetic structures~\cite{Chudnovsky1997}, exciton-polariton condensates~\cite{Alodjants2017}, etc. The approach is based on mapping the problem into the imaginary time domain, where the quantum tunneling is described as thermon oscillations in an inverse potential. We examine an effective potential for thermon quasiparticle oscillations in the conditions of the SC state formation. We establish significantly-quantum features of BJJ systems in Sec.~4 using the quasiprobability (Wigner) function analysis. We consider the limit, when the average particle number is large enough and the continuous variables approach is valid to the Wigner function in the particle number - phase difference space. Sec.~5 presents one of the main results of the paper; we discuss the quantum-classical crossover for the SC state formation problem at finite temperatures and weak tunneling dissipation, cf.~\cite{Caldeira,Hanggi1984,Larkin1983}. The numerical estimations support our results. The conclusion summarizes the results obtained.


\section{BJJ model}

\begin{figure}[h!]
\centerline{\includegraphics[width=0.75\linewidth]{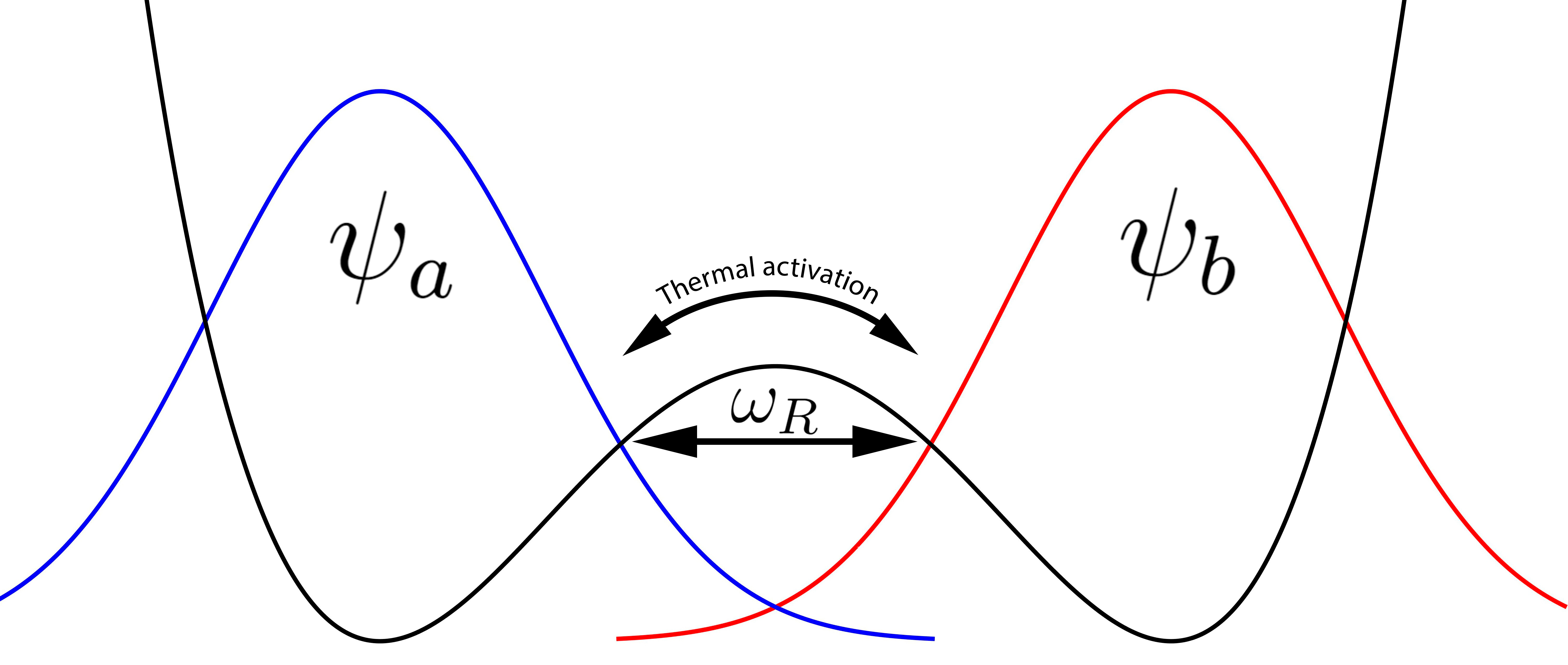}}
\caption{Scheme of the BJJ system consisting of double-well potential and the atom BECs. Wave functions $\psi_{a,b} = \sqrt{N_{a,b}}e^{i\theta_{a,b}}$ describe the envelopes of the BECs in semiclassical approximation; they overlap between the wells, which leads to tunnel (Josephson) coupling between the BECs with rate $\omega_R$. On the other hand, at finite temperatures, the atoms can hop above the potential well due to the thermal activation. $N_a$ and $N_b$ are average particle numbers in the wells, respectively; $\theta_a$ and $\theta_b$ are well-defined phases of wave functions in semiclassical approximation.} 
\label{FIG:Scheme}
\end{figure}

Consider two closely spaced atom BECs trapped in a double-well potential, see Figure~\ref{FIG:Scheme}. Such a system is known from literature as bosonic Josephson junctions (BJJ), see e.g.~\cite{Leggett2001,Pezze,Smerzi1997,Kohler,Rubeni} and references therein. In this paper, we restrict ourselves to two-mode approximation valid for the modest number of particles $N$ and some appropriate characteristic trap size, see e.g.\cite{Anglin}. 

In the second quantization form, a two-mode Hamiltonian of BJJ system looks like 
\begin{equation}\label{Ham1}
 \hat{H} = -\frac{u}{4}(\hat{a}^\dag\hat{a} - \hat{b}^\dag\hat{b})^2 - \frac{\hbar\omega_R}{2}(\hat{a}^\dag\hat{b} + \hat{b}^\dag\hat{a}),
\end{equation}
where $\hat{a}^\dag$ ($\hat{a}$) and $\hat{b}^\dag$ ($\hat{b}$) are creation (annihilation) operators of particles with mass $M$ to the left and right of the barrier, respectively, see Figure~\ref{FIG:Scheme}. In~\eqref{Ham1} $u = 4\pi\hbar^2|a_{sc}|/a_{\perp}^3M$ is the Kerr-like nonlinear interaction strength; $|a_{sc}|$ is the absolute value of atom-atom scattering length; $a_{\perp}$ is a characteristic spatial scale (trap size in the transverse direction); $\omega_R$ is the Rabi frequency. Notice, we consider attractively interacting particles with $a_{sc} < 0$ that possess $u > 0$: the sign of the term is explicitly accounted in~\eqref{Ham1}.

The ground state of the BJJ system in Fig.~\ref{FIG:Scheme} can be found in a two-mode Fock-state basis
\begin{equation}\label{Fock}
 |\Psi (t) \rangle = \sum_{n=0}^N A_n (t)|n,N-n\rangle,
\end{equation}
where $|n,N-n\rangle\equiv |n\rangle_a \otimes |N-n\rangle_b= |n_a\rangle \otimes |n_b\rangle$ is a tensor product of Fock states. The coefficients $A_n(t)$ in~\eqref{Fock} obey a usual normalization condition and may be obtained from the solution of the Schr{\"o}dinger equation $i\hbar\dot{A}_n (t) = \langle n,N-n| \hat{H} |\Psi(t) \rangle$ with Hamiltonian~\eqref{Ham1}; the dot denotes the derivative with respect to time. Finally, we obtain (cf.~\cite{Haigh2010})
\begin{equation}\label{master-equation}
 i\hbar \dot{A}_n = \alpha_n A_n + \beta_n A_{n+1} + \beta_{n-1} A_{n-1},
\end{equation}
where 
\begin{equation}\label{Ancf_1}
 \alpha_n = -\frac{\hbar\omega_R }{2} s \lambda\left(\frac{2n}{N} - 1 \right)^2,
\end{equation}
\begin{equation}\label{Ancf_2}
 \beta_n = -\frac{\hbar\omega_R}{2}\sqrt{(n+1)(N-n)},
\end{equation}
and
\begin{equation}\label{lambd}
 \lambda = \frac{2us}{\hbar\omega_R}\equiv\frac{uN}{\hbar\omega_R} 
\end{equation}
is a vital parameter that characterizes BJJ key features; $s\equiv N/2$.

\begin{figure}[t]
\begin{minipage}[h!]{0.49\linewidth}
\center{\includegraphics[width=\linewidth]{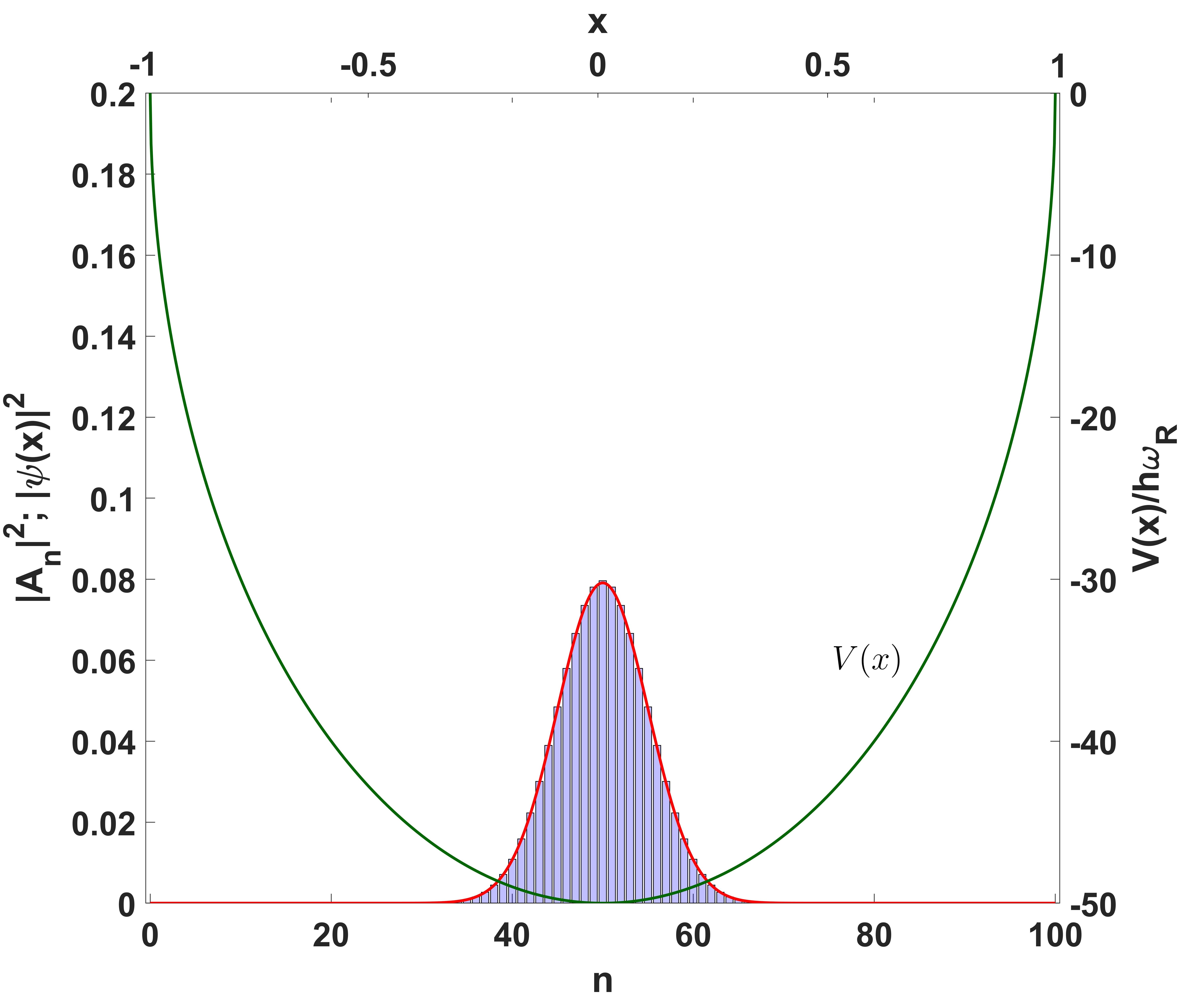} \\ a)}
\end{minipage}
\hfill
\begin{minipage}[h!]{0.49\linewidth}
\center{\includegraphics[width=\linewidth]{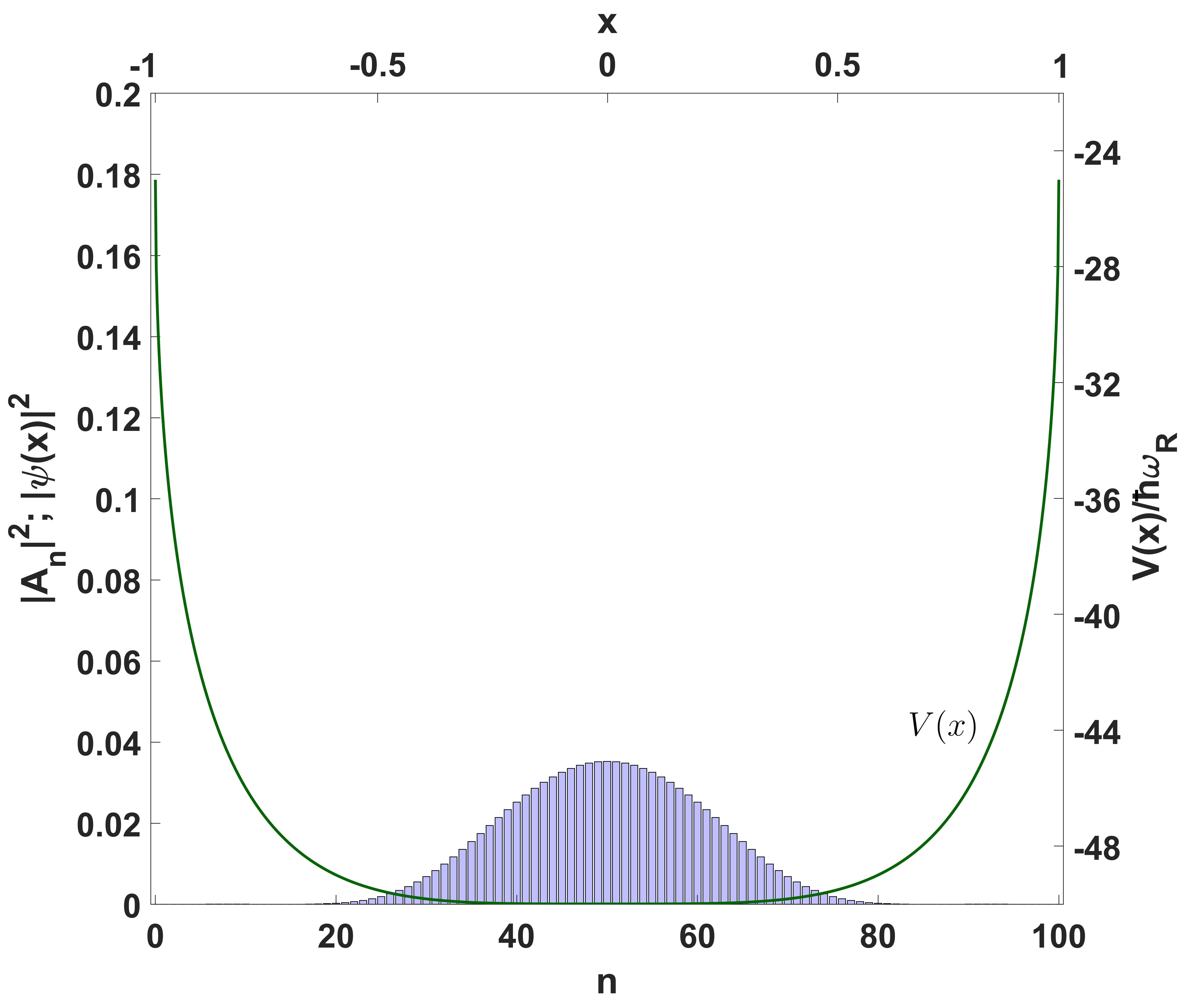} \\ b)}
\end{minipage}
\vfill
\begin{minipage}[h!]{0.49\linewidth}
\center{\includegraphics[width=\linewidth]{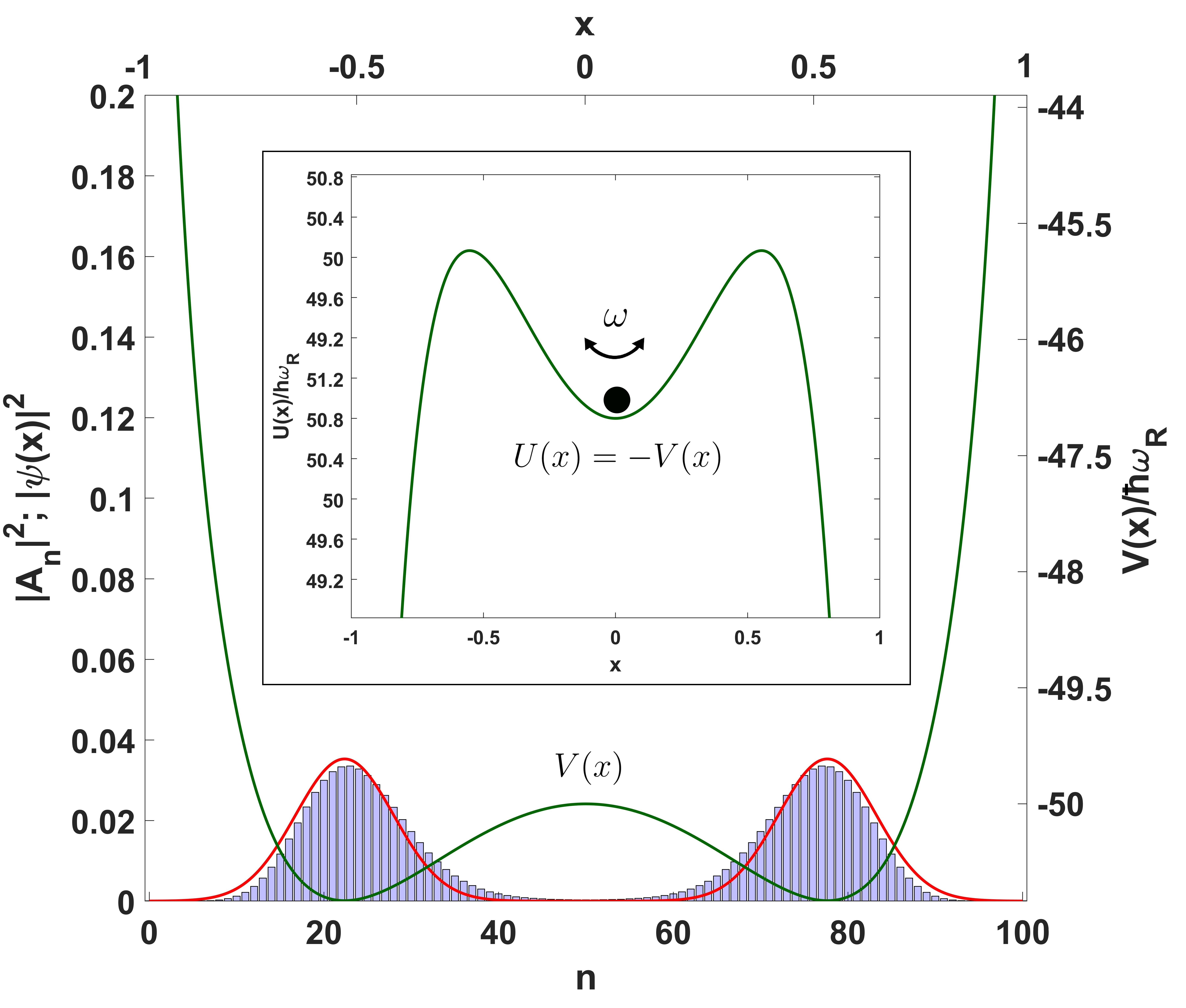} \\ c)}
\end{minipage}
\hfill
\begin{minipage}[h!]{0.49\linewidth}
\center{\includegraphics[width=\linewidth]{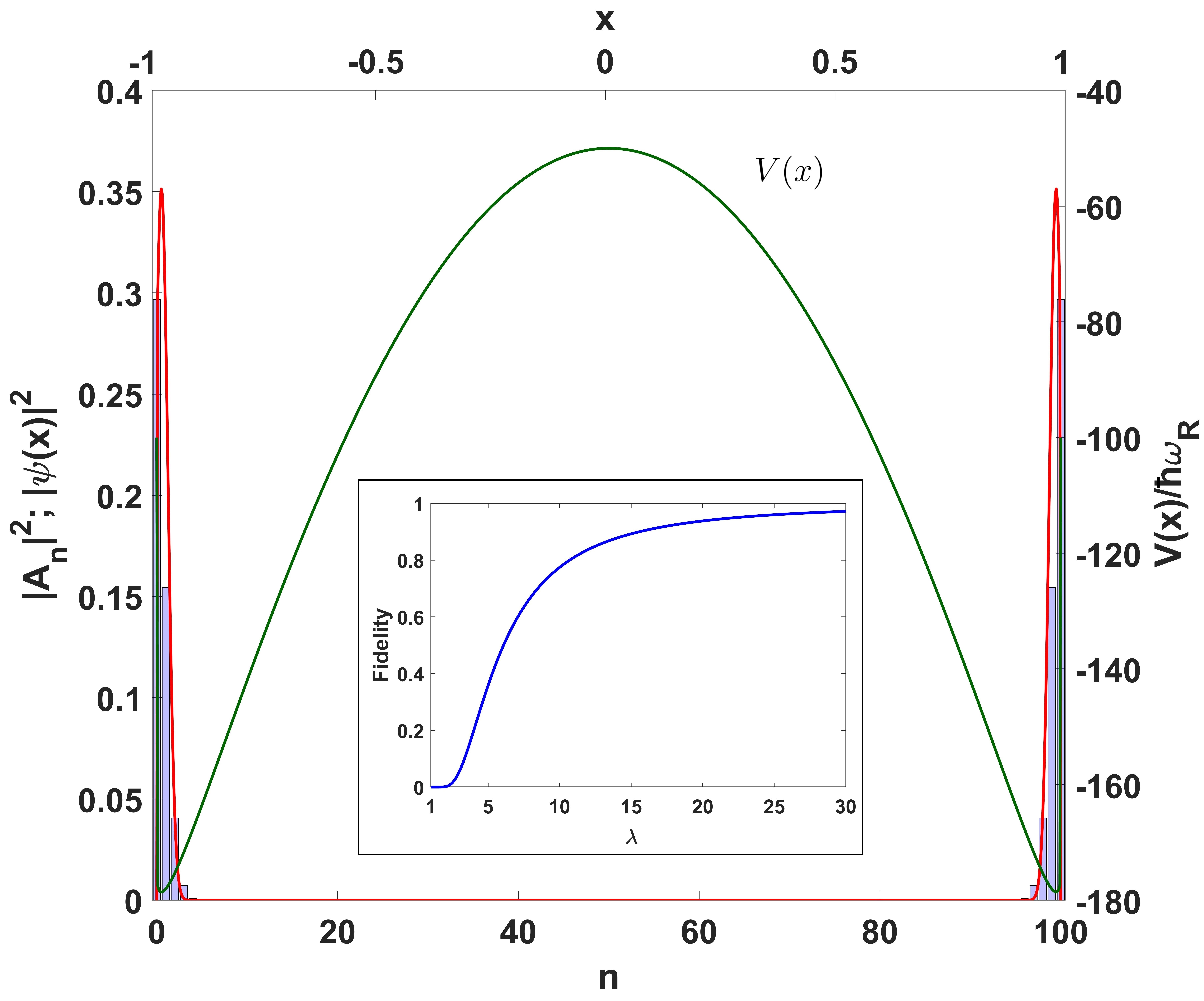} \\ d)}
\end{minipage}
\caption{BJJ ground state probabilities $|A_n|^2$ (the blue bars) vs. $n$ and approximate (Gaussian) envelopes $|\psi(x)|^2$ (the red curves) vs. $x = 2n/N - 1$ for (a) $\lambda = 0$ ($x_0 = 0$, $\sigma = 0.14$), (b) $\lambda = 1.0$, (c) $\lambda = 1.2$ ($x_0 = 0.55$, $\sigma = 0.16$), and (d) $\lambda = 7.0$ ($x_0 = 0.99$, $\sigma = 0.02$); the number of particles is $N = 100$. The green curves correspond to the effective potential $V(x)$ normalized on $\hbar\omega_R$, see Eq.~\eqref{pot1}. The relative error of Gaussian approximation is less than 4\% for (a), (c) and exceeds this value for (d). The inset in (c) demonstrates the small oscillations of thermon quasiparticle (shown in black) with frequency $\omega$ in effective inverse potential $U(x)=-V(x)$. The inset in (d) demonstrates the fidelity of the $N00N$ state formation in the BJJ system. For other details see the text.}
\label{FIG:Bars-n-pot}
\end{figure}

We are interested in stationary solutions of~\eqref{master-equation}, considering $A_n (t) = A_n e^{-i\frac{E}{\hbar}t}$, where $E$ is the energy of the BJJ ground state. We have discussed energy spectrum peculiarities for~\eqref{master-equation},~\eqref{Ancf_1} in~\cite{Tsarev2020}. In Fig.~\ref{FIG:Bars-n-pot} we represent ground state probability distributions $|A_n|^2$ (the blue bars) for various values of $\lambda$. For $\lambda<1$ and $N\gg1$ the initial distribution approaches the Gaussian one; in Fig.~\ref{FIG:Bars-n-pot}(a) such a distribution is shown by the red curve for non-interacting particles, $u=0$. Quantum phase transition to the SC state occurs at $\lambda=1$, cf.~\cite{Rubeni}. 

Two ``halves" of the SC state occur at $\lambda>1$. As $\lambda$ increases, they move from each other to the domain edges, $n=0,N$, increasing distinguishability of the SC state ``halves". Moreover, for $\lambda\gg1$ the SC state approaches the $N00N$ state:
 \begin{equation}\label{N00N} 
 |N00N\rangle = \frac{1}{\sqrt{2}}(|N,0\rangle + |0,N\rangle).
\end{equation}
We introduce fidelity $F$ of the $N00N$ state formation in the BJJ system in the form 
\begin{equation}\label{fidel} 
F = |\langle N00N|\Psi(\lambda)\rangle|^2.
\end{equation}

The inset in Fig.~\ref{FIG:Bars-n-pot}(d) clearly demonstrates the ideal (pure) $N00N$ state formation ($F=1$) in the limit of the infinitely large $\lambda$-parameter. For fidelity $F$ close to unity we can speak about the $N00N$-like state formation that, simply, represents a superposition of Fock states possessing two large components at the edges of particle number domain, $n=0$ and $n=N$, respectively, cf.~\cite{Tsarev2020}. The features of the $N00N$-like states with $\lambda\geq2$ and fidelity $F<1$, are close to the SC-states located at the edges of the discussed particle number domain. 
 

\section{Effective potential approach to the SC state}

We can utilize the effective potential method to characterize the obtained SC states if the total number of particles, $N$, is large enough, $N\gg1$. In this limit, we are able to introduce a continuous variable 
\begin{equation}\label{xa}
x=\frac{n_b-n_a}{N}=\frac{2n}{N}-1, 
\end{equation}
that characterizes normalized a particle number imbalance in potential wells. Variable $x$ varies within the window $-1\leq x \leq 1$; maximal, $x=1$, and minimal, $x=-1$, values of $x$ correspond to $n=N$ and $n=0$ number of particles in the ``$a$"-well, respectively. 

In the continuous model framework we replace $A_n$ with $\psi(x)$ assuming 
$A_{n+1}+A_{n-1}-2A_n \to \frac{1}{s^2} \frac{d^2}{dx^2} \psi(x)$. Eq.~\eqref{master-equation} now has a form 
\begin{equation}\label{diff1}
 \frac{\hbar\omega_R}{2s}\sqrt{1-x^2} \frac{d^2}{dx^2} \psi(x) + s\frac{\hbar\omega_R}{2}\left[\lambda x^2 + 2\sqrt{1-x^2} \right]\psi(x) = -E\psi(x).
\end{equation}
 
We recognize Eq.~\eqref{diff1} as a new Schr{\"o}dinger equation 
\begin{equation}\label{Harm_osc}
\frac{\hbar^2}{2m_{eff}} \frac{d^2}{dx^2} \psi(x) - \big(V(x) - E \big)\psi(x) = 0,
\end{equation}
that establishes mapping of the two-mode condensate to some fictitious macroscopic (FM) quantum particle with effective mass 
\begin{equation}\label{mass}
m_{eff} = \frac{s\hbar}{\omega_R\sqrt{(1 - x^2)}}\equiv \frac{m}{\sqrt{(1 - x^2)}}, 
\end{equation}
which moves in the effective potential (cf.~\cite{Alodjants2017}),
\begin{equation}\label{pot1}
V(x) = -s\frac{\hbar\omega_R}{2}\left(\lambda x^2 + 2\sqrt{1-x^2} \right).
\end{equation}

Notably, coefficient $\sqrt{1-x^2}$ in effective mass denominator~\eqref{mass} plays no essential role for further analysis of~\eqref{diff1},~\eqref{pot1}. 

Normalized potential $V(x)$ for various $\lambda$ is shown in Fig.~\ref{FIG:Bars-n-pot} by the green solid curves. As seen from Fig.~\ref{FIG:Bars-n-pot}(b-c), at $\lambda=1$ the potential changes its form from a single-well ($\lambda<1$) to the double-well ($\lambda>1$) structure, which indicates the formation of the Schr{\"o}dinger cat states in the potential minima.

At $\lambda<1$ potential $V(x)$ demonstrates a single minimum at $x = x_{min} = 0$, see e.g. Fig.~\ref{FIG:Bars-n-pot}(a). In the vicinity of this point the FM particle has effective mass $m_{eff} \simeq m =\hbar s/\omega_R$ and possesses Gaussian wave function 
\begin{equation}\label{Gauss}
 \psi(x) = \frac{1}{\sqrt{\sigma}\pi^{1/4}}\exp\left[{-\frac{x^2}{2\sigma^2}}\right],
 \end{equation}
where $\sigma$ is a width of the wave function. 

For $\lambda>1$ two minima with coordinates $x_{min,\pm} = \pm\sqrt{1 - \lambda^{-2}}$ occur for the effective potential $V(x)$. The FM particle has effective mass $m_{eff} = \lambda m$ in this case, see~\eqref{mass}. The SC state may be established as a superposition of two well-resolved Gaussian wave packets 
\begin{equation}\label{Gauss_b}
\psi(x) = C\left(\exp\left[{-\frac{(x-x_0)^2}{2\sigma^2}}\right] + \exp\left[{-\frac{(x+x_0)^2}{2\sigma^2}}\right]\right)
\end{equation}
located at two minima $x_{min,\pm}$ of potential $V(x)$ and separated by distance $d=2x_0 \equiv 2|x_{min}|$. In~\eqref{Gauss_b} $C = \Big (2\sigma\sqrt{\pi}\left(1 + \exp\left[-\frac{x_0^2}{\sigma^2}\right]\right)\Big)^{-1/2}$ is the normalization constant physically relevant to the amplitude of SC states. Notice, $\exp\left[-\frac{x_0^2}{\sigma^2}\right] \ll 1$ for $\lambda>1$, and, as a result, we can effectively neglect the overlapping of two terms assuming $x_0$ being large enough in comparison with $\sigma$ and $C\approx\frac{1}{\sqrt{2\sigma}\pi^{1/4}}$.

In the vicinity of the phase transition point, $\lambda = 1$, potential~\eqref{pot1} becomes essentially anharmonic, and Gaussian approximate solutions~\eqref{Gauss} and~\eqref{Gauss_b} are no longer valid.
 
It is useful to define the height of the potential barrier, $V_0\equiv V(0)-V(x_0)$, in Fig.~\ref{FIG:Bars-n-pot}(b,c) that occurs between two minima $x_{min,\pm}$ at $\lambda>1$; from~\eqref{pot1} we obtain 
\begin{equation}\label{pot21}
V_0 = s\frac{\hbar\omega_R}{2}\frac{\left(\lambda - 1\right)^2}{\lambda}
\end{equation}

To obtain the width of the wave packets, $\sigma$, in~\eqref{Gauss} and~\eqref{Gauss_b}, we expand~\eqref{pot1} in a Taylor series in the vicinity of $x_{min}$:
\begin{equation}\label{pot2}
 V(x) \simeq c_2(x - x_{min})^2,
\end{equation}
where $c_2$ is given in the form 
\begin{equation}\label{c20}
c_2 = s\frac{\hbar\omega_R}{2}\left(\left(1-x_{min}^2\right)^{-\frac{3}{2}} - \lambda\right). 
\end{equation}
In~\eqref{pot2} we omit the constant part of potential $V(x)$. Notice, this approach is not valid in the phase transition point, where $\lambda=1$, $x_{min}=0$, and $c_2=0$.

Eq.~\eqref{pot2} implies width of the ground state~\eqref{Gauss} in the form 
\begin{equation}\label{sigma}
\sigma = \frac{2m_{eff}}{\hbar^2}c_2^{-1/4}. 
\end{equation}
In particular, from~\eqref{sigma} it immediately follows that the width of the wave packets is $\sigma = (s^2(1 - \lambda))^{-\frac{1}{4}}$ for $\lambda<1$ and $\sigma = (s^2\lambda^2(\lambda^2 - 1))^{-\frac{1}{4}}$ for $\lambda>1$, respectively. In the limit of the $N00N$-like state formation, i.e. for large enough $\lambda$, the width of the wave packets approaches
\begin{equation}\label{sigmaN}
\sigma = \frac{1}{\sqrt{s}\lambda}. 
\end{equation}

The amplitude of the SC state $C^2 \simeq \frac{1}{2\sigma\sqrt{\pi}}$ may be obtained from Eqs. \eqref{Gauss_b}, \eqref{sigma} as
\begin{equation}\label{ampl}
C^2 = \frac{\sqrt{s\lambda}}{2\sqrt{\pi}}(\lambda^2 - 1)^{\frac{1}{4}}. 
\end{equation}
From \eqref{ampl} it follows that $C^2$ grows with $\lambda$. Moreover, in the limit of the $N00N$ state formation,  $\lambda\gg1$, the SC state amplitude grows approximately linearly with $\lambda$, i.e. $C^2 = \frac{\sqrt{s}}{2\sqrt{\pi}}\lambda\gg1$.

Thus, in this case we can obtain very narrow Gaussian wave packets separated by distance $d\approx 2\sqrt{1-\lambda^{-2}}$, see Fig.~\ref{FIG:Bars-n-pot}(d). In the limit of the pure $N00N$ state formation, $\sigma\to \infty$, $d=2$, and the Gaussian approach is not valid. In this limiting case, the effective potential approximation cannot be exploited. 

At $\lambda>1$, the condensate possesses small (harmonic) oscillations in the vicinity of potential $V(x)$ minima $x_{min,\pm}$ with angular frequency 
\begin{equation}\label{freq1}
\omega_0 =\sqrt{\frac{2c_2}{m_{eff}}}=\omega_R\sqrt{\lambda^2-1}. 
\end{equation}
In the limit of the $N00N$ state formation, i.e. for large enough $\lambda$, from~\eqref{freq1} we have $\omega_0 \approx \omega_R\lambda$. 


\section{Wigner function}

The non-classical behaviour of the SC states in the BJJ model may be elucidated by the Wigner function for conjugate (non-commuting) quantum variables~\cite{Hillery}. In the limit of the large particle number, $N$, the energy (phase) spectrum becomes quasi-continuous. In this limit, we can recognize them as continuous variables introducing Wigner function
\begin{equation}\label{Wigner}
W(x,p) = \frac{1}{\pi}\int_{-\infty}^\infty d\zeta \psi^*(x+\zeta)\psi(x-\zeta)e^{2ip\zeta},
\end{equation}
where $\psi$ is the wave function defined in~\eqref{Gauss}; $x$ and $p$ play the role of ``position" and ``momentum" variables associated with the particle number and phase $\theta = \theta_a - \theta_b$ difference variables, respectively, see Fig.~\ref{FIG:Scheme}. 
 
\begin{figure}[t]
\begin{minipage}[h!]{0.49\linewidth}
\center{\includegraphics[width=\linewidth]{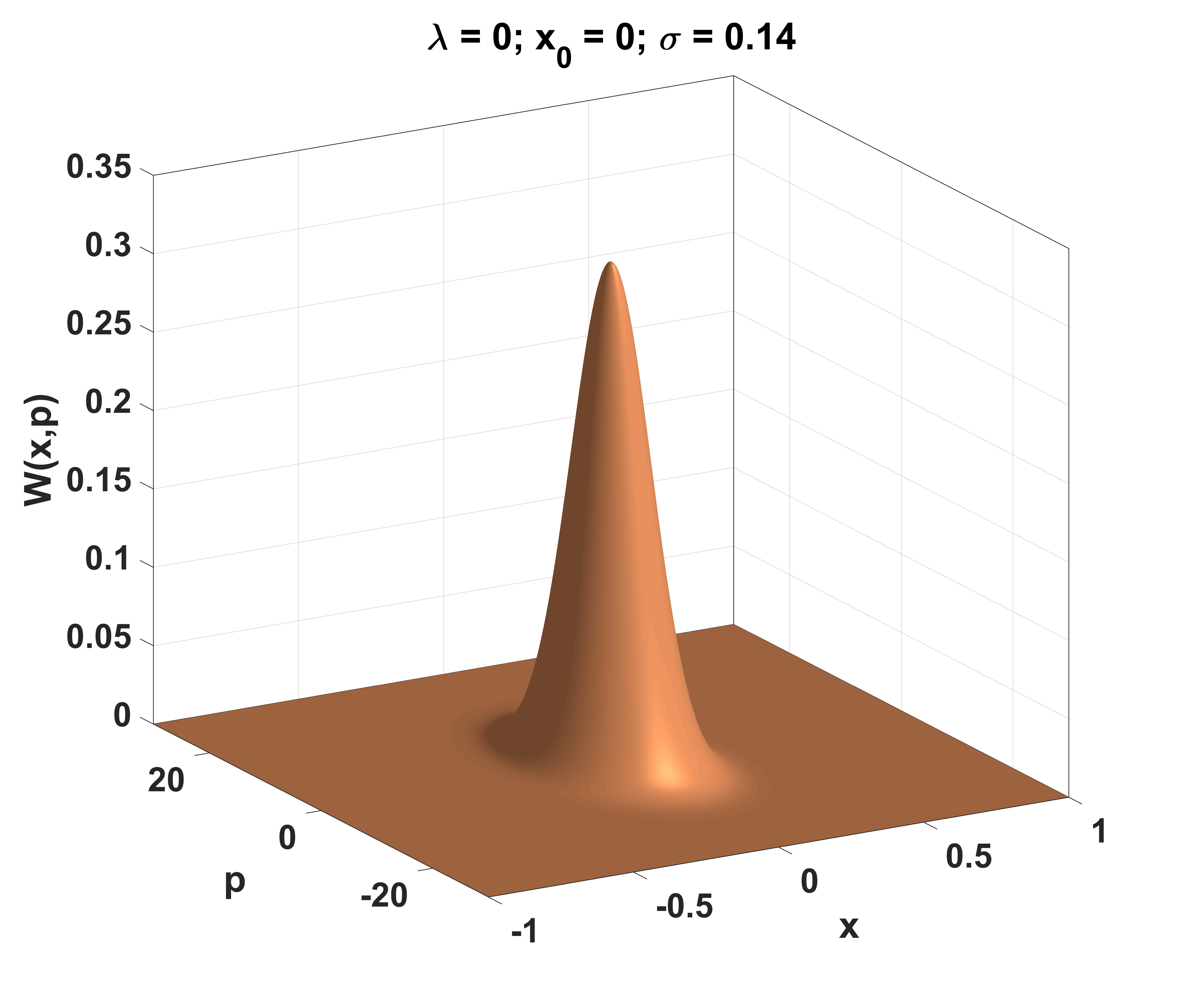} \\ a)}
\end{minipage}
\hfill
\begin{minipage}[h!]{0.49\linewidth}
\center{\includegraphics[width=\linewidth]{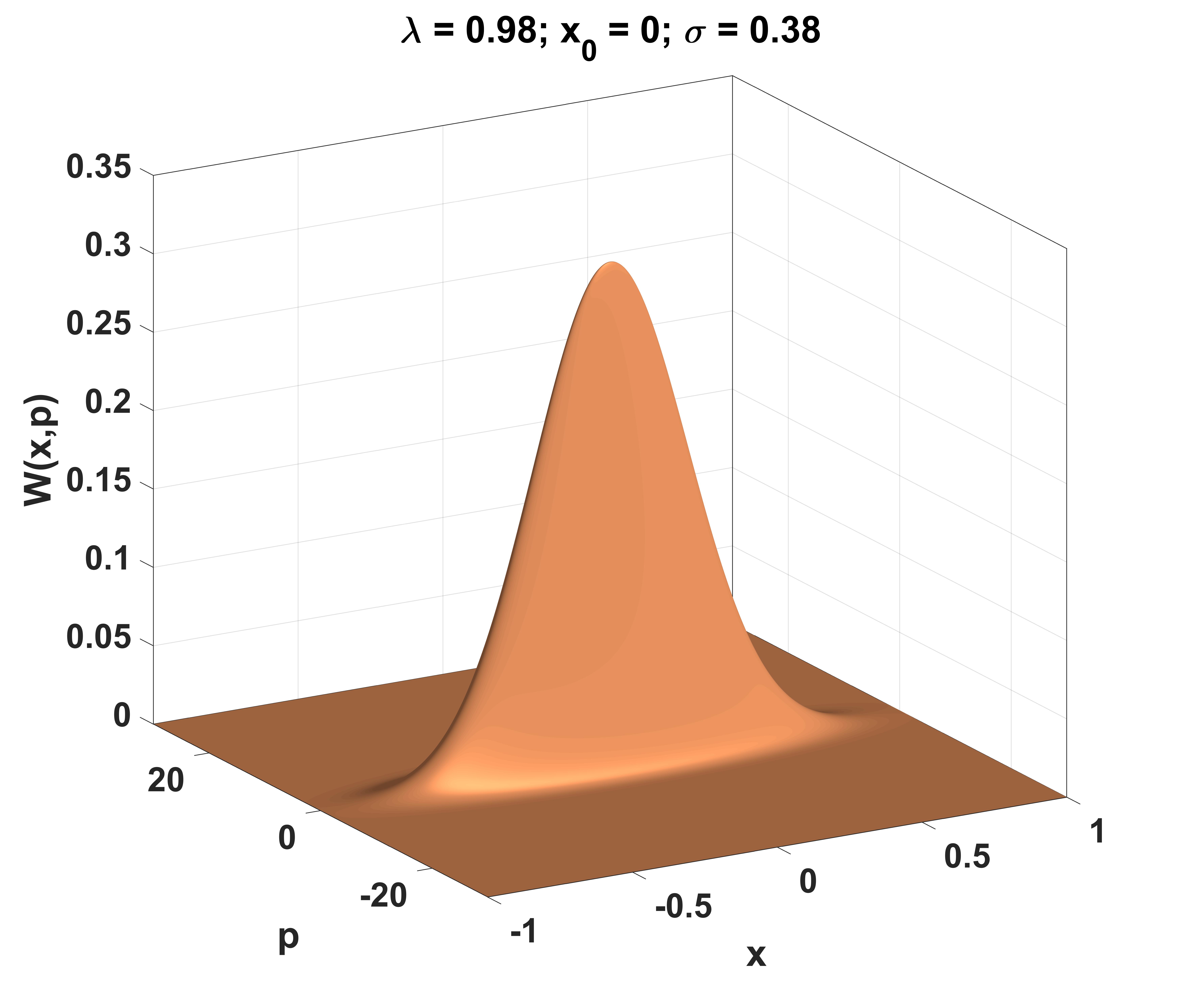} \\ b)}
\end{minipage}
\vfill
\begin{minipage}[h!]{0.49\linewidth}
\center{\includegraphics[width=\linewidth]{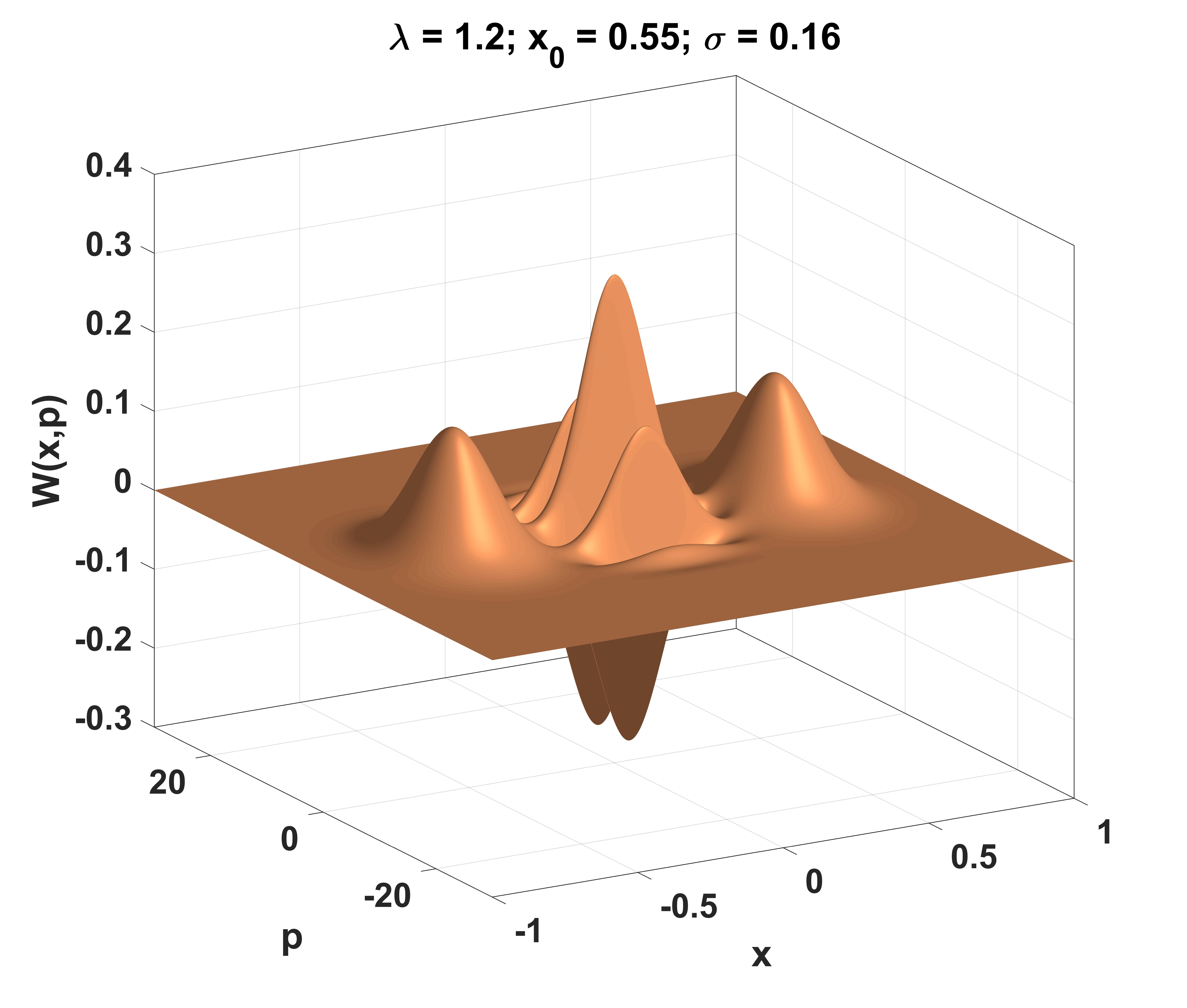} \\ c)}
\end{minipage}
\hfill
\begin{minipage}[h!]{0.49\linewidth}
\center{\includegraphics[width=\linewidth]{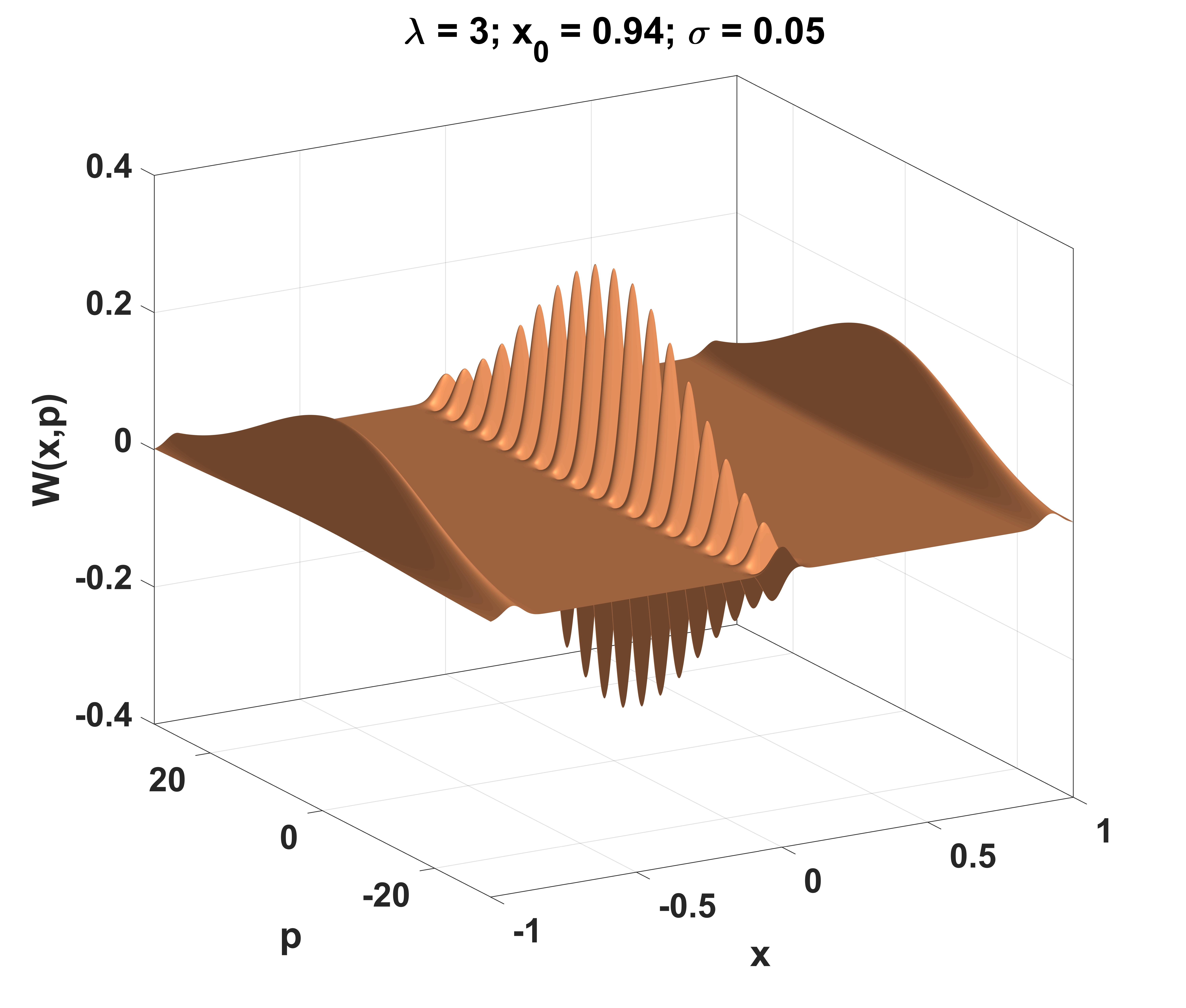} \\ d)}
\end{minipage}
\caption{3D plots of Wigner function $W(x,p)$ for vital parameter (a) $\lambda=0$, (b) $\lambda=0.98$, (c) $\lambda = 1.2$, and $\lambda = 3$. The total particle number in condensates is $N = 100$; other other parameters are represented in plots.}
\label{FIG:Wigner}
\end{figure} 
 
Substituting~\eqref{Gauss} into~\eqref{Wigner} at $\lambda<1$ one can obtain
\begin{equation}\label{Wigner1}
W(x,p) = \frac{1}{\pi}\exp\left[-\frac{x^2}{\sigma^2} - \sigma^2p^2\right]
\end{equation}
with $\sigma^{2}=1/s\sqrt{1-\lambda}$. In Fig.~\ref{FIG:Wigner}(a) we represent Wigner function~\eqref{Wigner1} for non-interacting particles ($\lambda=0$) that corresponds to Gaussian state in Fig.~\ref{FIG:Bars-n-pot}(a). In the phase transition point the Wigner function becomes essentially non-Gaussian; in Fig.~\ref{FIG:Wigner}(b) we demonstrate the behaviour of the Wigner function in the vicinity of this point. 

Similarly, for $\lambda>1$ we obtain 
\begin{eqnarray}\label{Wigner2}
W(x,p)=\frac{1}{2\pi}\exp\left[-\frac{(x+x_0)^2}{\sigma^2}\right]\exp\left[-\sigma^2p^2\right] +\\ 
+\frac{1}{2\pi}\exp\left[-\frac{(x-x_0)^2}{\sigma^2}\right]\exp\left[-\sigma^2p^2\right]+\frac{1}{\pi}\exp\left[-\frac{x^2}{\sigma^2} - \sigma^2p^2\right]\cos\left[2px_0\right], \nonumber
\end{eqnarray}
where $x_0 = \sqrt{1 - \lambda^{-2}}$ and $\sigma^2 = 1/s\lambda\sqrt{\lambda^2-1}$.

The first two terms in~\eqref{Wigner2} describe the Gaussian peaks (cat states) at $x\approx\pm x_0$ in the framework of the Wigner function $W(x,p)$ approach. The third term in~\eqref{Wigner2} is relevant to the quantum interference and characterizes oscillations in $p$-axis at $x\approx0$, see Fig.~\ref{FIG:Wigner}(c). With increasing the $\lambda$-parameter these peaks move toward the edges of the $x$-domain establishing the $N00N$-like states; in Fig.~\ref{FIG:Wigner}(d) we represent the Wigner function for the SC states located closely to the edges $x=\pm1$ at $\lambda=3$. Notice, the analytical approach to the Wigner function established in~\eqref{Wigner2} fails with the further increase of the $\lambda$-parameter. Moreover, for $\lambda\gg1$, i.e. in the limit of the pure $N00N$ state formation, we must account the discretization of the particle number - phase difference space, cf.~\cite{Mukunda}. 


\section{Discussion}

The analysis performed is valid in the zero-temperature limit, when the quantum tunneling between the junctions represents the primary physical process responsible for the SC state formation. In this limit, the BJJ system may be suitable for applications in quantum annealing and other approaches to quantum computing, cf.~\cite{Alodjants2017}. However, at finite temperatures, in the presence of coupling with the environment, the quantum tunneling may be suppressed by the thermal activation process, which supports classical bistable system occurring on the basis of the BJJ, instead of the SC state. Thus, it is important to elucidate the conditions, under which the SC states can be obtained in the real-world experiments operating with various BJJ systems. Here, we shortly discuss such conditions accounting dissipation effects appearing at finite temperatures and our results obtained in~\cite{Alodjants2017}. 

The role of finite temperatures may be accounted by the ``bounce" technique originally proposed in~\cite{Langer} and then developed in~\cite{Larkin1983,Callan,Melnikov,Hanggi1985,Grabert1987} in the framework of the metastable state decay problem stated at finite temperatures in the presence (or without) dissipation. In particular, this technique presumes utilization of the imaginary time variable, $t = -i\hbar/k_BT$, that allows mapping of quantum particle moving in the double-well potential to so-called, thermon quasiparticle, which harmonically oscillates within the inverted potential with the period, $\tau\simeq 1/T$, where $T$ is the temperature. 

To be more specific, for our problem in Fig.~\ref{FIG:Bars-n-pot} we examine the inverse potential, $U(x) = -V(x)$, which in the vicinity of saddle point $x\approx 0$ has a form (cf.~\eqref{pot1}) 
\begin{equation}\label{inv_potentia} 
U(x) \simeq U(0) - \left(c_2x^2 + c_4x^4\right) + O(x^6),
\end{equation}
where coefficients $c_{2,4}$ are 
\begin{equation}\label{c2} 
c_2 = - s\frac{\hbar\omega_R}{2}(\lambda - 1);
\end{equation}
\begin{equation}\label{c4} 
c_4 = \frac{s\hbar\omega_R}{8}.
\end{equation}
Formally, within the harmonic oscillations approximation, Eq.~\eqref{c2} establishes the same coefficient $c_2$, represented in~\eqref{c20} with $x_{min}=0$.

In the presence of the thermally assisted effects, there exists critical temperature $T_c$ that corresponds to the phase transition from the classical activation to the quantum tunneling phenomenon. In particular, the critical temperature, 
\begin{equation}\label{temp1} 
 T_c =\frac{\hbar\omega}{2\pi k_B},
\end{equation}
is determined through motion of the thermon quasiparticle within the inverted potential, $U(x)$, near its bottom with angular frequency $\omega$ defined as (see the insert in Fig.~\ref{FIG:Bars-n-pot}(c)) 
\begin{equation}\label{freq} 
\omega = \sqrt{\frac{U''_{xx}(0)}{m}} = \omega_R\sqrt{\lambda-1}=\frac{\omega_0}{\sqrt{\lambda+1}}, 
\end{equation}
where $U''_{xx}(0)$ denotes the second derivative of the inverse potential calculated at the saddle point $x=0$ for $\lambda>1$, cf.~\cite{Grabert1987}.

Inverse potential $U(x)$ with coefficients $c_2$ and $c_4$ given in~\eqref{inv_potentia} - \eqref{c4} specifies the second-order phase transition from the classical to quantum features for the BJJ system, cf.~\cite{Chudnovsky1997,Alodjants2017}. Noteworthy, in the vicinity of $\lambda=1$, the frequency of the thermon oscillations, $\omega$, and the critical temperature, $T_c$, go to zero. Such a behaviour manifests a purely quantum nature of the phase transition to the SC state that occurs in the BJJ system at $\lambda=1$. 

At high enough temperatures, $T\gg T_c$, classical thermal effects prevail, and the BJJ system undergoes thermal activation. In particular, in this limit FM particle escapes with the rate
\begin{equation}\label{Aren} 
 \Gamma_{cl} = \frac{\omega_0}{2\pi}e^{-V_0/k_BT}, 
\end{equation}
that resembles the famous Arrhenius law; $V_0$ is the height of the potential barrier formed by effective potential $V(x)$ at $\lambda >1$ and specified in~\eqref{pot21}. Typically, weak metastability of the decayed quantum state requires some conditions for the barrier height, $V_0$, that should obey inequalities
\begin{equation}\label{cond} 
\frac{V_0}{k_B T}\gg 1, \quad \frac{V_0}{\hbar \omega_0} = \frac{s}{2\lambda}\frac{(\lambda - 1)^{3/2}}{\sqrt{\lambda+1}}\gg 1. 
\end{equation}
Conditions~\eqref{cond} are easily fulfilled for the large number of particles, $N=2s\gg1$.

In the presence of weak dissipation, critical temperature $T_c$ in~\eqref{temp1} modifies and may be represented as (cf.~\cite{Grabert1987})
\begin{equation}\label{dissipation} 
 T_c = \frac{\hbar \omega \alpha}{2\pi k_B}, 
\end{equation}
where $\alpha=\sqrt{1+(\gamma/2\omega)^2}-\gamma/2\omega$ is the factor that depends on (Ohmic) damping rate $\gamma$ and characterizes the slowing down of the FM particle dynamics caused by friction; $\omega$ is defined in~\eqref{freq}. Here, we focus on the underdamping regime when $\gamma\ll\omega$. In particular, Eq.~\eqref{Aren} with the dissipation escape rate has a form
\begin{equation}\label{preexp} 
\Gamma = \frac{\omega_0\alpha}{2\pi} f_q e^{-V_0/k_BT}, 
\end{equation}
where we introduce the dimensionless quantum correction factor, $f_q$, that characterizes the deviation of escape rate $\Gamma$ from the classical one, $\Gamma_{cl}$. It is possible to show (see e.g.~\cite{Hanggi1985}) that the quantum corrections well above the crossover temperature $T_c$ lead to $f_q$ in the form 
\begin{equation}\label{f} 
f_q \simeq \textrm{exp}\Big[\frac{\hbar^2(\omega_0^2+\omega^2 )}{24(k_B T)^2}\Big]. 
\end{equation}

Eq.~\eqref{f} exhibits the improvement of the escape rate due to quantum tunneling effects. Notice, the exponent degree in~\eqref{f} does not depend on dissipation rate $\gamma$. 

In the low temperature limit, for $T \ll T_c$, Eqs.~\eqref{Aren}, \eqref{preexp} are not applicable. At $T\to 0$, we deal with purely quantum tunneling effects characterized by the tunneling rate, 
\begin{equation}\label{Kram} 
 \Gamma_{q} \simeq e^{-B}, 
\end{equation}
where exponent degree $B$ is independent on temperature $T$. 

At temperatures $T\leq T_c$, the escape rate predominantly is characterized by quantum tunneling. In the crossover region, close to the critical value, $T_c$, the interplay occurs between the thermal and quantum fluctuations. The exact equations for the escape rate in this case are cumbersome, cf.~\cite{Grabert1987}, and useless for numerical estimations. However, we can estimate the exponent degree, $B_c$, when quantum fluctuations become crucially important. From~\eqref{dissipation}, \eqref{preexp}, and~\eqref{Kram} we can obtain 
\begin{equation}\label{B} 
B_c\simeq \frac{V_0}{k_B T_c} = \frac{s \pi}{\alpha\lambda}\left(\lambda-1\right)^{\frac{3}{2}}.
\end{equation}
Thus, the SC state observation, which is based on the essentially quantum tunneling processes, may be obtained for the barriers possessing $B\geq B_c$. 

Consider some numerical estimations for the discussed characteristic parameters of the BJJ system. To be more specific, we limit ourselves to two-mode condensate of ${}^7$Li atoms (the mass of the atom is $M = 1.165\times10^{-26}$ kg) possessing negative scattering length $a_{sc} = -0.21$ nm tuned via the Feshbach resonance technique~\cite{Khaykovich}. Let $uN/k_B = 10$ nK for $N = N_0=100$ particles. This value is taken sufficiently below the critical number of particles required for condensate to collapse. In particular, we suppose the atoms to be in a harmonic trap with $\omega_\perp = 2\pi\times 967$ Hz, which gives $a_\perp = 1.22$ $\mu$m. Then, for $\omega_R = 2\pi\times208$ Hz we can tune the BJJ system nearby the critical value, $\lambda = 1$. For $N<N_0$, the BJJ system is prepared in an atom-coherent state. On the other hand, at $N>N_0$ we can observe the BJJ SC state; in the limit $N\gg N_0$ the $N00N$-like state is formed. 
 
Substituting $\omega_R$ into Eq.~\eqref{freq}, we obtain critical temperature $T_c = T_0\sqrt{\lambda - 1}$, where $T_0 = \hbar\omega_R/2\pi k_B = 1.6$ nK. For comparison, the critical temperature of Lithium Bose condensation is typically $T_{BEC}\simeq100$ nK~\cite{Parkins1998,Bradley1995,Bradley1997}, and the inequality $T_0\ll T_{BEC}$ is typically fulfilled. Thus, we can conclude that the SC and $N00N$-like states for atomic condensate systems may be obtained under significantly low temperatures obeying the $T\ll T_0\ll T_{BEC}$ condition. 


\section{Conclusion}

In this work, we have examined a seminal BJJ quantum system that provides the SC and $N00N$ states formation depending on vital parameter $\lambda$, which represents the medium Kerr-like nonlinearity parameter ratio to the tunneling (Josephson coupling) rate. We have explored the effective potential approach that is clear from Fig.~\ref{FIG:Bars-n-pot} and posed the accurate analytical description of the BJJ system behaviour for large number of particles. Such a description presumes the Wigner function approach, which is plotted in Fig.~\ref{FIG:Wigner} and, together with Fig.~\ref{FIG:Bars-n-pot}, reveals the main features of the BJJ system. In particular, we have shown that quantum phase transition to the SC state occurs at $\lambda=1$ in the zero temperature limit. Below this point, i.e. for small $\lambda$, the system may be described by the atom-coherent state that approaches the Gaussian wave function. Nearby the phase transition point, $\lambda=1$, the effective potential is anharmonic, and the wave function of the BJJ system is significantly non-Gaussian. However, at $\lambda>1$, the SC states manifest non-classical properties of the BJJ device. For large enough $\lambda$, the separation (in the phase space) between the SC state ``halves" grows and the $N00N$-like states form in the BJJ system. The ideal $N00N$ state appears in the limit of infinitely large $\lambda$ that actually requires tremendous number of particles. 

We have examined the possibility to attain the SC state at finite temperatures and weak dissipation. They both lead to the critical temperature, $T_c$, that defines the characteristic scale, when quantum tunneling is suppressed, and the thermal activation effects become important; it happens at $T>T_c$. The transition from the classical hopping to quantum tunneling is a second-order phase transition occurring at critical temperature $T_c$. We anticipate the SC and $N00N$-like states formation sufficiently below $T_c$. Our numerical estimations demonstrate that $T_c$ lies below $T_{BEC}$ - the characteristic atomic condensation temperature. 

The arguments presented above can make certain difficulties in the experiment with the SC and, especially, $N00N$-like states design. Practically, tremendous values of the $\lambda$-parameter requiring large enough number of particles look non-realistic. In particular, the number of particles for atoms possessing negative scattering length is still limited by the condensate collapsing phenomenon. At the same time, the two-mode condensate approach violates for $N\simeq 1000$ and more. The spatial degree of freedom cannot be neglected in this case. In this sense, it is instructive to consider the soliton Josephson junction system that we suggested in~\cite{Tsarev2020} and examined in~\cite{Alodjants2022}. Such a system demonstrates the $N00N$-like state formation at moderate values of the crucial $\lambda$-parameter. We plan to represent our results in forthcoming publications. 


\section{Acknowledgments}

The results of this work related to the SC state description at zero temperatures by the effective potential method are given in Secs.~2,3 and were funded by Ministry of Science and Higher Education of the Russian Federation and South Ural State University (agreement №075-15-2022-1116). Studies of the Wigner function for the SC and $N00N$ states, finite temperature results described in Secs. 4,5 are supported by research project no. 2019-1442 of Ministry of Science and Higher Education of the Russian Federation.


\end{document}